\documentclass[         %
aps,                    
prd,                    
showpacs,               
superscriptaddress,    
nofootinbib,            
showkeys,               %
preprintnumbers,        %
floatfix]               
{revtex4}               
\usepackage{graphicx,longtable,amsmath}

\numberwithin{equation}{section}

\begin{document}
\title{Solar Neutrino Matter Effects Redux}
\author{        A.~B. Balantekin}
\email{         baha@physics.wisc.edu}
\author{A.~Malkus}
\email{acmalkus@ncsu.edu}
\altaffiliation{Current address: Department of Physics, North Carolina State University, Raleigh, NC 27695 USA.}
\affiliation{Department of Physics, University of Wisconsin - Madison, Wisconsin 53706 USA }

\date{\today}
\begin{abstract}

Following recent low-threshold analysis of the Sudbury Neutrino Observatory and asymmetry measurements of the BOREXINO Collaboration of the solar neutrino flux, we revisit the analysis of the matter effects in the Sun. We show that solar neutrino data constrains the mixing angle $\theta_{13}$ poorly and that subdominant Standard Model effects can mimic the effects of the physics beyond the Standard Model.   

\end{abstract}
\medskip
\preprint{}
\maketitle

\vskip 1.3cm

\section{Introduction}
\label{Section: Introduction}

Over the last two decades, solar neutrino physics evolved from a qualitative description of the data into a precision science. Following the seminal experiment of Davis \cite{Cleveland:1994er}, experiments were performed to measure the predominant pp neutrinos \cite{Abdurashitov:1999zd,Altmann:2000ft}, as well as the higher energy neutrinos: Borexino measured the line flux coming from $^7$Be neutrinos \cite{Arpesella:2008mt}. The continuous flux of the $^8$B neutrinos were measured at SuperKamiokande (SuperK) \cite{Fukuda:2001nj} and at Sudbury Neutrino Observatory (SNO) 
\cite{Ahmad:2001an} with a recent confirmation from the KamLAND experiment \cite{Abe:2011em}. 
The Cerenkov detector experiments SuperK and SNO are high statistics experiments. In addition, SNO experiment was successful in measuring the total high-energy solar neutrino flux using two different methods to detect the neutral current breakup of deuteron \cite{Aharmim:2008kc}. This measurement,  providing the total flux independent of neutrino mixing and matter effects, is in general agreement with the predictions of the Standard Solar Model \cite{Bahcall:2004pz,TurckChieze:2001ye}.  In parallel with these experimental developments there have been increasingly refined analyses of the solar neutrino data alone or combined with the data from accelerator, reactor, and atmospheric neutrino measurements (see e.g. Refs. \cite{Fogli:2008jx,Fogli:2005cq,GonzalezGarcia:2010er,Fogli:2011qn,Balantekin:2003dc}). In the particular case of solar neutrinos, two distinct regions of the neutrino parameter space are identified: the so-called LOW region corresponding to smaller values of $\delta m_{21}^2$ ($\sim 10^{-7}$ eV$^2$) and the so-called LMA region\footnote{LMA stands for "large mixing angle". This is somewhat of a misnomer as the LOW region is also operative for large mixing angles.} corresponding to larger values of $\delta m_{21}^2$ ($\sim 10^{-5}$ eV$^2$). The LMA region was originally adopted as the appropriate region of the parameters space because of the KamLAND reactor neutrino observations. 

The original analyses of the various phases of the SNO experiment were performed with a rather large energy threshold. More recently SNO was able to lower its threshold to an effective electron kinetic energy of 3.5 MeV \cite{Aharmim:2009gd}. With such high precision reached in detecting solar neutrino flux and associated observables, it is worthwhile to visit various matter effects and their ramifications. Furthermore, our measurements of the solar neutrino flux are not complete: neutrinos from the CN cycle 
\cite{Adelberger:2010qa} are not yet identified. (These neutrinos could probe the interaction between the early Sun  and its protoplanetary disk \cite{Serenelli:2011py}). The sun can also be used as a laboratory for searching for physics both beyond the standard solar model and the Standard Model of the particle physics. Examples of the former includes searching for effects of density fluctuations 
\cite{Balantekin:2003qm,Fogli:2007tx}. Examples of the latter include the effects of large neutrino magnetic moments \cite{Lim:1987tk}, neutrino interactions beyond those in he Standard Model such as flavor-changing neutral currents \cite{Friedland:2004pp}, mass-varying neutrinos \cite{Cirelli:2005sg}, and new long-range forces \cite{GonzalezGarcia:2006vp}. A search for the subdominant physics operational in the Sun requires a thorough understanding of the dominant matter and neutrino mixing effects. For example as we demonstrate below, the high-statistics solar neutrino experiments SNO and SuperK cannot really distinguish between LOW and LMA solutions. Breaking this degeneracy requires using the KamLAND data. To completely rule out the unlikely possibility of CPT violation (i.e the possibility of $\delta m^2$ being different for solar {\it neutrinos} and the reactor {\it antineutrinos}), one has to  examine another matter-effect, namely the day-night asymmetry. Indeed Borexino experiment recently measured this asymmetry for the $^7$Be neutrinos, verifying the LMA solution (and supporting the CPT invariance) using only solar neutrinos \cite{Bellini:2011yj}. 

In the next section we provide a critical look at the SNO low-threshold analysis and Borexino Day-Night Asymmetry measurement. We show that the solar neutrino data constrains the mixing angle 
$\theta_{13}$ poorly. In section III, we examine the effects of new physics beyond the Standard Model and show that subdominant Standard Model effects can mimic new physics. (Mathematical details are included in an Appendix). Finally, a brief discussion of our results conclude the paper.

\section{SNO low-threshold analysis and Borexino Day-Night Asymmetry measurement}

\begin{figure}[t]
\begin{center}
\includegraphics[scale=0.3]{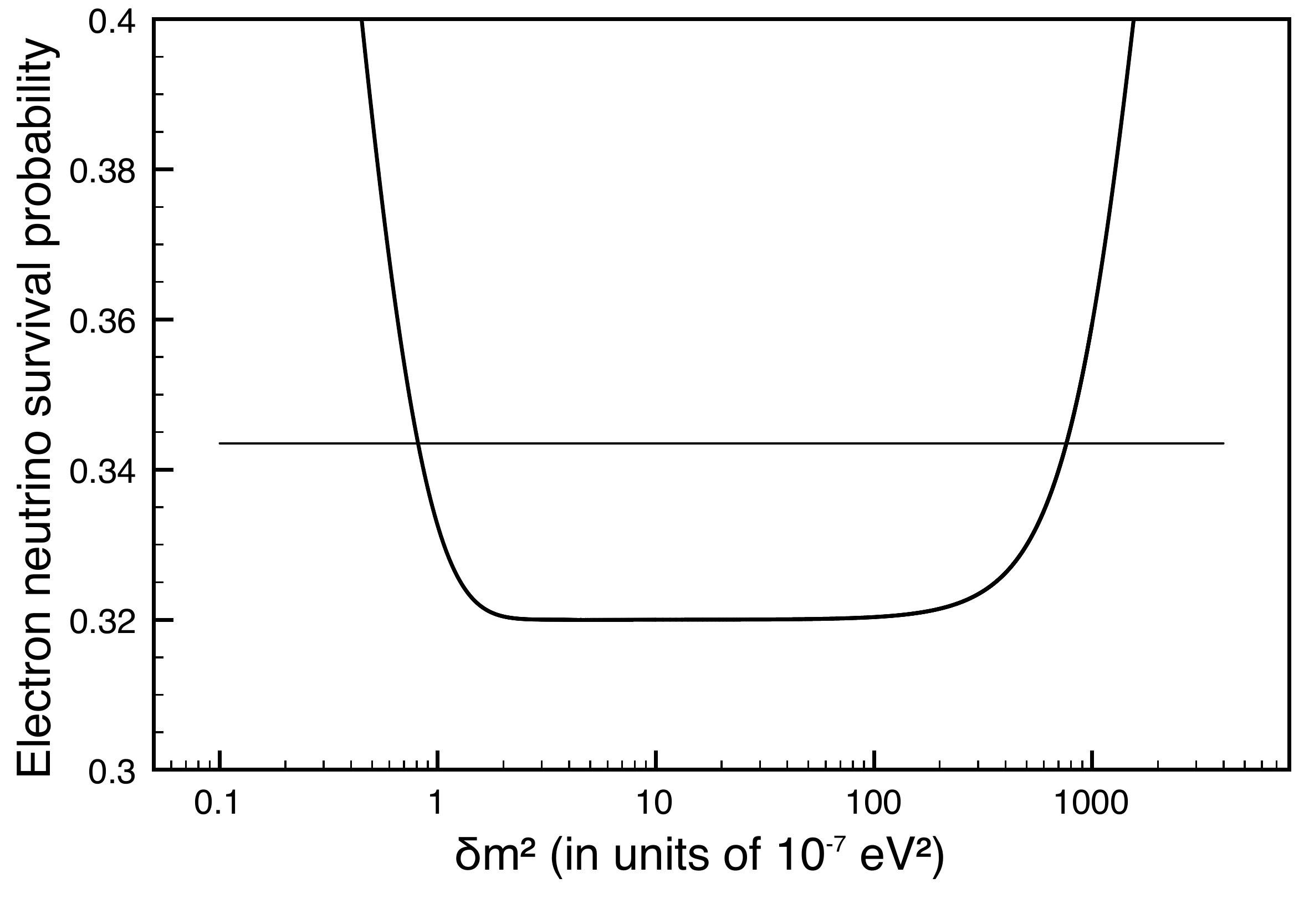}
\caption{Solar electron neutrino survival probability as a function of $\delta m^2$ with $\cos 2 \theta_v = 
0.36$ and $E_{\nu} = 10$ MeV. The average solar electron density at the neutrino production region is taken to be $N_e$ = 100 $N_A$/cm$^3$. The horizontal line depicts the value measured by SNO at the low-threshold analysis \cite{Aharmim:2009gd}.}
\label{Fig:1}
\end{center}
\end{figure}

Low-energy threshold analysis of the phase I and phase II data sets of the Sudbury Neutrino Observatory was also reported as the electron-neutrino survival probability expanded around the neutrino energy $E_{\nu} = 10$ MeV \cite{Aharmim:2009gd}. In this section we first show that, with most mixing angles, the solar electron-neutrino survival probability is doubly degenerate in 
$\delta m^2_{21}$.  Assuming two-flavor mixing, the electron neutrino survival probability detected on the Earth during the day and  averaged over the Earth-Sun distance is \cite{Balantekin:1998yb} 
\begin{eqnarray}
\label{1}
P( \nu_e \rightarrow \nu_e) &=& \frac{1}{2} +\frac{1}{2} \cos 2 \theta_v \langle \cos 2 \theta_i 
\rangle_{source}  \left( 1 - 2 |\Psi_{2,(S)}|^2 \right) \nonumber \\
&-& \frac{1}{2}  \cos 2 \theta_v \langle \sin 2 \theta_i \rangle_{source} \left( \Psi_{1,(S)} \Psi_{2,(S)} + 
 \Psi^*_{1,(S)} \Psi^*_{2,(S)} \right) ,
\end{eqnarray}
where $\theta_v$ and $\theta_i$ are the vacuum and initial matter mixing angles respectively, $\langle \cdots \rangle_{source}$ denotes averaging over the neutrino production region in the Sun,  and 
$\Psi_{1,(S)}$  and $\Psi_{2,(S)}$ are the matter eigenstates calculated on the surface of the Sun. The quantity $|\Psi_{2,(S)}|^2$ is sometimes referred to as the hopping probability, $P_{hop}$. In the regions of the neutrino parameter space consistent with the solar neutrino observation one has either $\cos 2 \theta_i \sim -1$ (the so-called LOW region) or $\Psi_{2,(S)} \sim 0$ (the so-called LMA , or large mixing angle, region). In both cases the last term in Eq. (\ref{1}) can be dropped and one obtains \cite{Haxton:1986dm} 
\begin{equation}
\label{2}
P^D( \nu_e \rightarrow \nu_e) = \frac{1}{2} +\frac{1}{2} \cos 2 \theta_v \langle \cos 2 \theta_i 
\rangle_{source}  \left( 1 - 2 P_{hop} \right) .
\end{equation}
For a given value of the mixing angle, the energy (or more correctly $\delta m^2/E$) dependence of the survival probability comes from the product $\cos 2 \theta_i  (1 - 2 P_{hop})$. For smaller values of $\delta m^2$ ($\sim 10^{-7}$ eV$^2$), even though the matter angle is at its maximal value, $P_{hop}$ is finite (but small) for a range of neutrino energies. For larger values of $\delta m^2$ ($\sim 10^{-5}$ eV$^2$), $P_{hop}$ vanishes, but the matter angle is no longer maximal, hence slowly changes with neutrino energy. Hence for a fixed value of the vacuum mixing angle, there are two values of $\delta m^2$ to obtain a given solar electron neutrino survival probability as depicted in Figure \ref{Fig:1}. 

\begin{figure}[t]
\begin{center}
\includegraphics[scale=0.3]{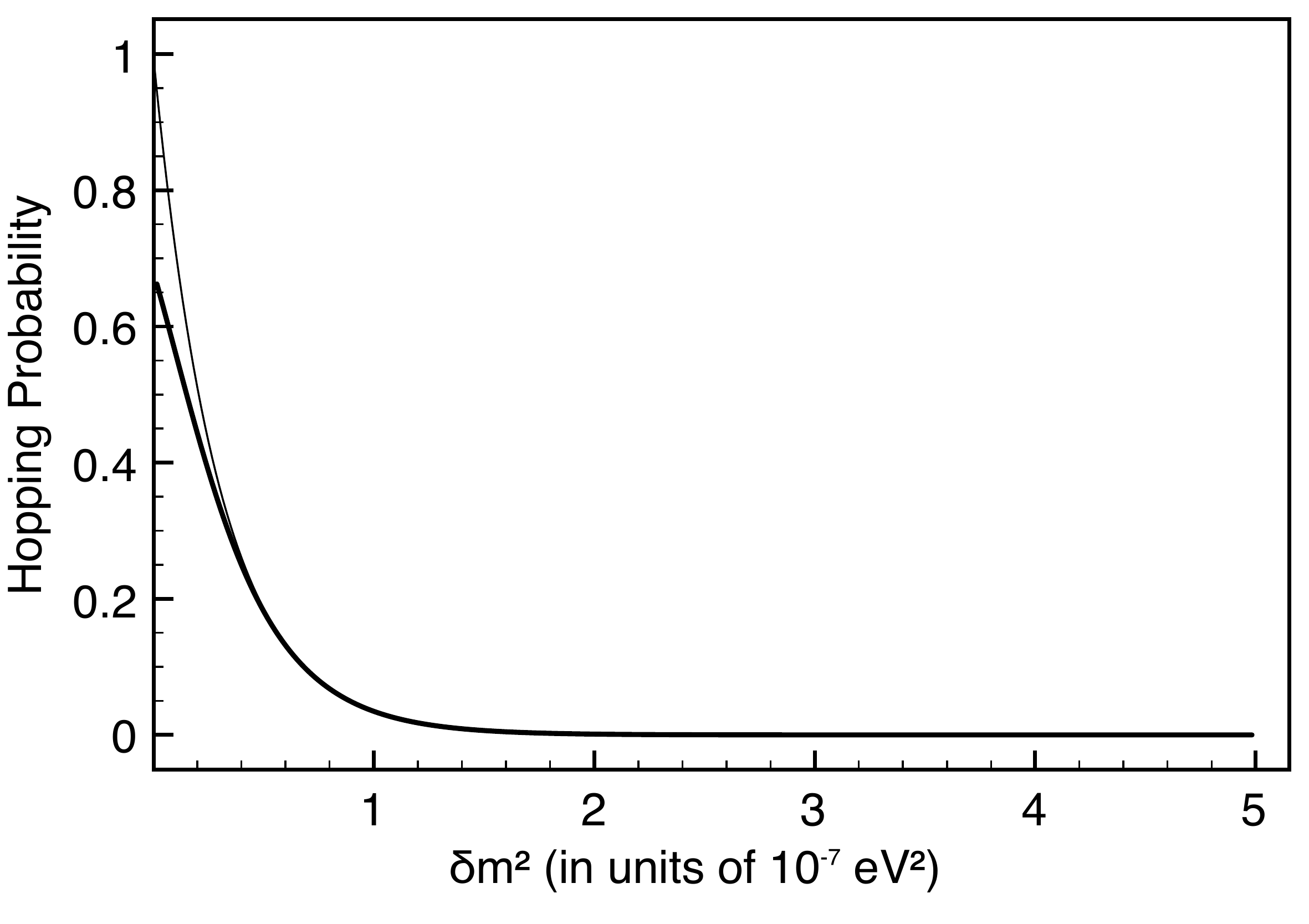}
\caption{A comparison of the hopping probabilities calculated for an exponential electron density. The exact result given by Eq. (\ref{6}) (thick line) is plotted versus the approximate result given by Eq. (\ref{7}) (thin line).The same neutrino parameters are used as in Fig. \ref{Fig:1}.}
\label{Fig:2}
\end{center}
\end{figure}

The cosine of the matter mixing angle is
\begin{equation}
\label{3}
\cos 2 \theta_M = - \frac{(2 B E - \cos 2 \theta_v)}{\sqrt{(2 B E - \cos 2 \theta_v)^2 + \sin^2 2 \theta_v}}
\end{equation}
where, for later convenience, we defined
\begin{equation}
\label{4}
B = \sqrt{2} G_F N_e / \delta m^2. 
\end{equation}
The Taylor expansion of Eq. (\ref{3}) around $E_0 = 10$ MeV gives
\begin{equation}
\label{5}
\cos 2 \theta_M = \cos 2 \theta_0 - 2B \frac{\sin^3 2 \theta_0}{\sin 2 \theta_v} (E - E_0) - 6 B^2 
\frac{\sin^4 2 \theta_0}{\sin^2 2 \theta_v} \cos 2 \theta_0 (E - E_0)^2 + \cdots .
\end{equation}
In the equations throughout  this section the subscript $0$ denotes a quantity calculated at the neutrino energy of 10 MeV. 
An approximate expression for the hopping probability can be estimated by exploiting the fact that the solar density decreases more or less exponentially, $N_e = N_0 \exp(-r/r_0)$. For such a density profile the hopping probability is given by \cite{Bruggen:1995hr}
\begin{equation}
\label{6}
P_{hop} = \frac{ e^{-\pi \delta (1-\cos 2 \theta_v)} - e^{-2\pi \delta}}{1 - e^{-2 \pi \delta}}
\end{equation}
where $\delta = r_0 \delta m^2/2E$. In the Sun, the arguments of the exponentials of Eq. (\ref{6}) are large and sometimes the approximate expression \cite{Pizzochero:1987fj}
\begin{equation}
\label{7}
P_{hop} \sim e^{-\pi \delta (1-\cos 2 \theta_v)} 
\end{equation}
is used. Hopping probabilities form these two expressions are plotted in Fig. \ref{Fig:2}. 
One sees that, for the pertinent values of $\delta m^2$, the difference between the hopping probabilities predicted by these two expressions are very small. For the arguments below we will use Eq. (\ref{7}). 
(Since the solar electron density is not exactly an exponential, it is also possible to utilize a generalization of Eq. (\ref{7}) to near exponential  densities \cite{Balantekin:1997fr} in the analysis  presented below).  Expanding the hopping probability of Eq. (\ref{7}), we get
\begin{equation}
\label{8}
P_{hop} = P_{hop}^0  - P_{hop}^0 \log P_{hop}^0 \left( \frac{E-E_0}{E_0} \right) 
+ \left( P_{hop}^0 \log P_{hop}^0 + \frac{1}{2} P_{hop}^0 (\log P_{hop}^0)^2 \right) \left( \frac{E - E_0}{E_0} \right)^2 + \cdots
\end{equation}

\begin{figure}[t]
\begin{center}
\includegraphics[scale=0.3]{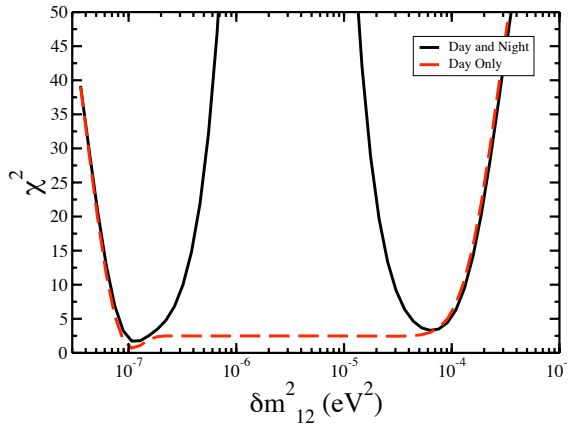}
\caption{Chi-square values of our numerical fit to the SNO LTE results with $\theta_{13} =0$. As this figure illustrates, it is not possible to distinguish LOW and LMA regions with the SNO data set}
\label{Fig:3}
\end{center}
\end{figure}

Low-energy threshold analysis of SNO reports the daytime solar electron neutrino survival probability  as \cite{Aharmim:2009gd}
\begin{equation}
\label{9}
P^D (\nu_e \rightarrow \nu_e) = c_0 + c_1(E - 10) + c_2 (E-10)^2, 
\end{equation}
where energies are measured in MeV. Note that the values of $c_1$ and $c_2$ given by SNO are consistent with zero.  In this regard, including a third order term in Eq. (\ref{9}) would have been impractical as it would have had a large uncertainty. From Eqs. (\ref{2}), (\ref{5}) and (\ref{8}) we expect 
\begin{equation}
\label{10}
\frac{2 c_0 - 1}{\cos 2 \theta_v} = \left\{
\begin{array}{rl}
 \cos 2 \theta_0 &  \text{for LMA}   \\
 (2P_{hop}^0 - 1)  & \text{for LOW}
\end{array}
\right. .
\end{equation}
Assuming the values $\cos 2 \theta_v = 0.36$, $\delta m_{21}^2 = 7.6 \times 10^{-5}$ eV$^2$ currently recommended by the Particle Data Group \cite{Nakamura:2010zzi}, and using the SNO central value of $c_0 = 0.3435$ , this gives $\cos 2 \theta_0 = -0.87$ which corresponds to an average initial density of $N_e \sim 100 \> N_A/cm^3$ in the production region of 10 MeV $^8$B neutrinos, in excellent agreement with the Standard Solar Model \cite{Bahcall:2004pz,TurckChieze:2001ye}. 
For the LOW solution we get a hopping probability of $P_{hop}^0 \sim 0.065$, yielding $\delta = 1.36$ or 
for $\delta_{21}^2 \sim 8.2 \times 10^{-8}$ eV$^2$, we get $r_0 = R_{\odot}/10.6$, again in good agreement with an exponential fit to the solar electron density profile \cite{bahcall}. 
For the parameter $c_1$ we expect
\begin{equation}
\label{11}
\frac{2 c_1}{\cos 2 \theta_v} = \left\{
\begin{array}{rl}
 - 2 B \frac{\sin^3 2 \theta_0}{\sin 2 \theta_v} &  \text{for LMA}   \\
 - P_{hop}^0 \log P_{hop} / E_0  & \text{for LOW}
\end{array}
\right. .
\end{equation} 
It is widely accepted that studies of the matter effects in solar neutrino oscillations over the last decade
established $\delta m_{21}^2$ to be positive \cite{GonzalezGarcia:2010er} yielding a positive value for $B$. 
This makes the sign of $c_1$ negative for the LMA solution in Eq. (\ref{11} in contrast with the SNO value. On the other hand the LOW solution has the same sign as the SNO value (note that $P_{hop} < 1$). 
These considerations suggest that SNO data will have a slight preference for the LOW solution. 
We present  numerically calculated $\chi^2$ for the entire SNO data set in Fig. \ref{Fig:3}. 
Clearly qualitative arguments given in this section are verified. 

\begin{figure}[t]
\begin{center}
\includegraphics[scale=0.3]{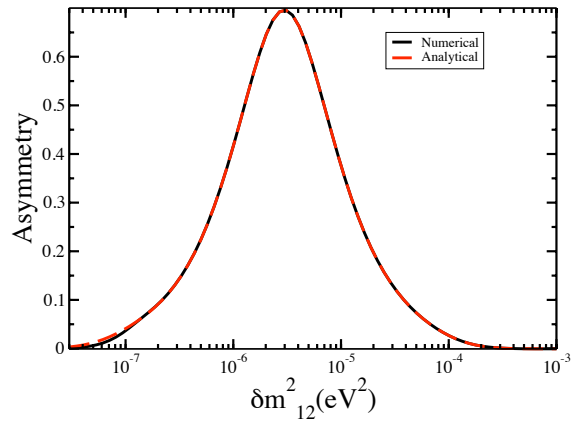}
\caption{Calculated day-night asymmetry for the SNO data set. The dashed line is for the exact numerical calculation and the solid line is given by the approximate analytical expression, Eq. (\ref{14}). }
\label{Fig:4}
\end{center}
\end{figure}

SNO low-energy threshold analysis also provides a fit to the day-night asymmetry. For the LMA and LOW regions night-time solar electron neutrino survival probability, averaged over the Earth-Sun distance is given by
\cite{GonzalezGarcia:2000ve,Minakata:2010be}:
\begin{equation}
\label{11a}
P^N( \nu_e \rightarrow \nu_e) = \frac{1}{2} +\frac{1}{2} \langle \cos 2 \theta_i \rangle_{source}  \left( 1 - 2 P_{hop} \right) \cos 2 \theta_e \cos 2(\theta_e - \theta_v) 
\end{equation}
where $\theta_e$ is the matter mixing angle inside the Earth. In writing Eq. (\ref{11}), matter density in the mantle of the Earth (through which neutrinos detected at SNO would travel) is assumed to be constant. 
The necessary expression for the day-night asymmetry is 
\begin{equation}
\label{14}
\frac{A}{2} = \frac{P^N-P^D}{P^N + P^D}. 
\end{equation}
Using the central value of the asymmetry extracted by SNO at $E_{\nu} = 10$ MeV, $A = 0.0325$, and 
the initial matter angle $\cos 2 \theta_i = - 0.87$ in the LMA region or the hopping probability $P_{hop} = 0.065$ in the LOW region, obtained above, 
we find that the degeneracy of the solution with respect to $\delta m_{21}^2$ survives both with day and night data. 
Note that the asymmetry quoted by SNO is consistent with zero when the statistical and systematic errors are taken into account.  
We also present numerically calculated day-night asymmetry in Fig. \ref{Fig:4}. 
Numerical results agree with the simple analytical expression, Eq. (\ref{14}) and the qualitative discussion presented above. 

\begin{figure}[b]
\begin{center}
\includegraphics[scale=0.3]{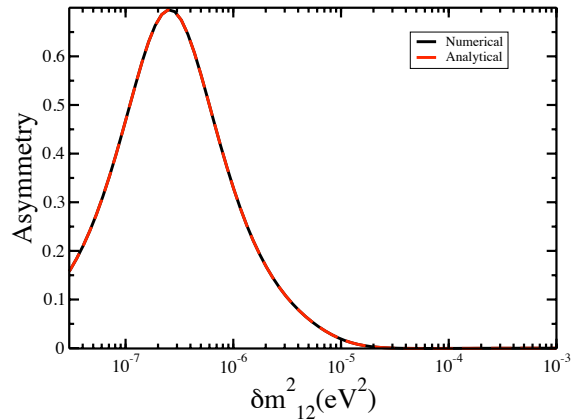}
\caption{Calculated day-night asymmetry for the Borexino data. The dashed line is for the exact numerical calculation and the solid line is given by the approximate analytical expression, Eq. (\ref{14}). }
\label{Fig:5}
\end{center}
\end{figure}

Recently the Borexino experiment reported a measurement of the day-night asymmetry for the $E_{\nu} = 0.86$ MeV $^7$Be line neutrinos \cite{Bellini:2011yj}. 
We show the calculated value of this asymmetry in Fig. \ref{Fig:5}. 
Clearly the reported value of A = 0.001 $\pm$ 0.012 (stat) $\pm$ 0.007 (syst) completely rules out the LOW region of the neutrino parameter space using only the solar {\it neutrino}  data without any direct reference to the reactor {\it antineutrinos}. 
This experimentally establishes that the neutrino and antineutrino mixings are indeed identical. 

So far we have been discussing only two-flavor mixing, which is only correct for $\theta_{13}=0$ 
\cite{Balantekin:1999dx}. Already several years ago it was pointed out that the tension between solar neutrino experiments and the KamLAND reactor neutrino experiment could be explained by a small, but non-zero value of $\theta_{13}$ \cite{Balantekin:2008zm}. 
SNO low-threshold analysis reports a value of $\theta_{13} = 7.2 ^{+2.0}_{-2.8}$ degrees, albeit with non-Gaussian errors \cite{Aharmim:2009gd}. A global analysis by the KamLAND  collaboration indicates $\sin^2 \theta_{13} = 0.009^{+0.013}_{-0.007}$ at the 79\% C.L. 
\cite{Gando:2010aa}. Best current indication of a non-zero value of $\theta_{13}$ are given by the T2K and MINOS experiments. T2K collaboration announced a 2.5$\sigma$ result: $0.03 < \sin^2 2\theta_{13} < 0.28$ for the normal and  $0.04 < \sin^2 2\theta_{13} < 0.34$ for the inverted hierarchies, both for $\delta_{CP}=0$ and maximal $\theta_{23}$ \cite{Abe:2011sj}. MINOS collaboration reports 
$0 < 2 \sin^2 2\theta_{13} \sin^2 \theta_{23}  < 0.12$ for the normal hierarchy and 
$0 < 2 \sin^2 2\theta_{13} \sin^2 \theta_{23}  < 0.19$ for the inverted hierarchy, again both for 
$\delta_{CP}=0$ \cite{minos}\footnote{One should note that {\em appearance} experiments can only measure the combination $ \sin^2 2\theta_{13} \sin^2 \theta_{23}$, hence the uncertainty in the measured value of $\theta_{23}$ needs to be folded in to quote a separate value of $\theta_{13}$. }.  Recent global analyses are consistent with these values 
\cite{Fogli:2011qn,Schwetz:2011zk}. 
Currently three reactor neutrino experiments (Double Chooz \cite{Ardellier:2006mn}, Daya Bay \cite{Guo:2007ug}, and RENO \cite{Ahn:2010vy}) are planning to directly measure this angle. 

To discuss the effects of  $\theta_{13}$ on solar neutrino data many times the formula 
\cite{Fogli:2001wi,Balantekin:2003dc,Fogli:2008jx}
\begin{equation}
\label{b1}
P_{3\times3}( \nu_e \rightarrow  \nu_e) = \cos^4{\theta_{13}} \> 
P_{2\times2}( \nu_e \rightarrow  \nu_e \>{\rm with}\> N_e
\cos^2{\theta_{13}})  + \sin^4{\theta_{13}} 
\end{equation}
is used. A discussion of the accuracy of this formula was given in Refs. \cite{Fogli:2001wi} and 
\cite{Balantekin:2011ta}. We will utilize this formula in the discussion given below. 

\begin{figure}[t]
\begin{center}
\includegraphics[scale=0.4]{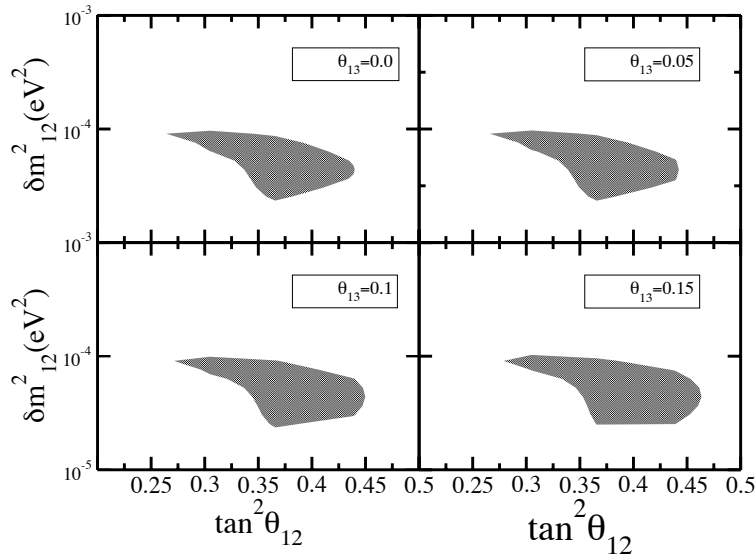}
\caption{The 90\% confidence level regions of the neutrino parameters allowed by only all the solar neutrino experiments for different values of $\theta_{13}$ (in radians).}
\label{Fig:6}
\end{center}
\end{figure}

\begin{figure}[t]
\begin{center}
\includegraphics[scale=0.3]{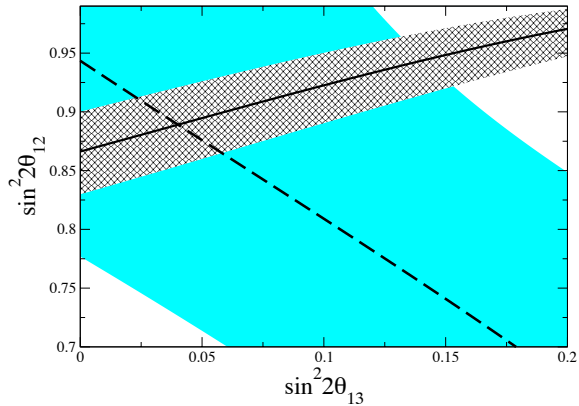}
\caption{Constant electron neutrino survival probability contours for a fixed value of  $\delta m^2_{12}$ (taken to be $7.6 \times 10^{-5}$ eV$^2$)  in the $\theta_{12}-\theta_{13}$ plane. The solid line is for a 10 MeV neutrino with survival probability set to the value given by SNO in the low-threshold analysis. The dashed line is for a 862 keV neutrino with survival probability set to the value given by the Borexino experiment. The wide shaded and hatched areas correspond to one-sigma uncertainties quoted by the Borexino and SNO experiments, respectively.}
\label{Fig:7}
\end{center}
\end{figure}

We present 90\% confidence level regions of the neutrino parameters allowed {\em only} by all the solar neutrino experiments for different values of $\theta_{13}$ in Fig. \ref{Fig:6}. 
An examination of the figure reveals that changing the value of $\theta_{13}$ has very little effect on the $\delta m^2_{21}$ and $\sin \theta_{12}$ parameter space.  
Reasons beyond the insensitivity of the solar neutrino data to the value of $\theta_{13}$ can be explicated by investigating constant $P(\nu_e \rightarrow \nu_e)$ contours in the $\theta_{12}-\theta_{13}$ plane. 
We present constant electron neutrino survival probability contour for a 10 MeV solar neutrino with $\delta m^2_{12} = 7.6 \times 10^{-5}$ eV$^2$  and a constant survival probability for a 0.862 MeV neutrino in the $\theta_{12}-\theta_{13}$ plane in Fig. \ref{Fig:7}. 
The survival probability for the 10 MeV neutrino is set to the value given by SNO in the low-threshold analysis and the probability for the 0.862 MeV neutrino is set to the value given by Borexino 
\cite{Bellini:2011rx}. 
In Ref. \cite{Goswami:2004cn} it was pointed out that the central values from low energy neutrino experiments and high energy neutrino experiments together close in on a small range of $\theta_{13}$.
However, the large uncertainties from the low energy experiment mean that even though the two isocontours intersect, the range of allowable values for $\theta_{13}$ is large. 
We conclude that the current solar neutrino data alone would constrain the mixing angle $\theta_{13}$ rather poorly. 

\section{Standard Model versus new physics in matter effects} 

A convenient way to parameterize the effect of new interactions on neutrinos traveling through media 
is   \cite{Friedland:2004pp,deGouvea:2004va,Huber:2002bi}
\begin{equation}
\label{deviation}
i \frac{\partial}{\partial t} 
\left(
\begin{array}{c}
  \Psi_e   \\
  \Psi_x
\end{array}
\right)
= \left[   \frac{\delta m^2_{21}}{4E}  \left(
\begin{array}{cc}
- \cos 2 \theta_{12}  &  \sin 2 \theta_{12}   \\
\sin 2 \theta_{12}  &   \cos 2 \theta_{12}
\end{array}
\right)
+ \frac{G_F N_e}{\sqrt{2}} 
\left(
\begin{array}{ccc}
 1+ \epsilon_{11} &   \epsilon_{12}^*  \\
\epsilon_{12}  & -1 - \epsilon_{11}     
\end{array}
\right)
\right] 
\left(
\begin{array}{c}
  \Psi_e   \\
  \Psi_x
\end{array}
\right)
\end{equation}
where the terms $\epsilon_{11}$ and $\epsilon_{12}$ represent contributions from new physics. Here we wish to emphasize that non-zero values of these terms do not necessarily imply new physics. (Clearly 
$\epsilon_{11} = -1$ and $\epsilon_{12}=0$ represent neutrino oscillations in vacuum). A derivation of the Eq. (\ref{b1}) is outlined in the Appendix. In solving the neutrino propagation with $\theta_{13} \neq 0$ one solves an effective two-flavor equation where $N_e$ is multiplied by $\cos^2 \theta_{13}$, which corresponds to $\epsilon_{11} = - \sin^2 \theta_{13}$ and $\epsilon_{12}=0$. The fact that various oscillation effects may be confused with non-standard interactions was already emphasized in Ref. 
\cite{Palazzo:2009rb}. 

We rewrite Eq. (\ref{loop}) for a neutral medium as 
\begin{equation}
\label{loop1}
V_{\tau \mu} =  N_e \tilde{V}_{\tau \mu}
\end{equation}
with 
\begin{equation}
\label{loop2}
\tilde{V}_{\tau \mu} = -  \frac{3 \sqrt{2} G_F\alpha}{\pi Y_e \sin^2 \theta_W}  \left( \frac{m_{\tau}}{m_W}\right)^2
\left\{ \log \frac{m_{\tau}}{m_W} + \left( \frac{2+Y_e}{6} \right) \right\} , 
\end{equation}
where the electron fraction, $Y_e$, for a neutral medium is given as
\begin{equation}
\label{ye}
Y_e = \frac{N_e}{N_p +N_n} = \frac{N_p}{N_p +N_n}. 
\end{equation}
Hence inclusion of the loop corrections would appear in the effective two-flavor evolution as 
\begin{equation}
\epsilon_{11} = - \sin^2 \theta_{13} + \tilde{V}_{\tau \mu} \left( \sin^2 \theta_{13} \cos^2 \theta_{23} - 
    2 \sin^2 \theta_{23} \right), 
\end{equation}
and 
\begin{equation}
\epsilon_{12} =  \sin \theta_{13} \sin 2 \theta_{23}  \tilde{V}_{\tau \mu} e^{-i \delta_{CP}} . 
\end{equation}

\begin{figure}[t]
\begin{center}
\includegraphics[scale=0.3]{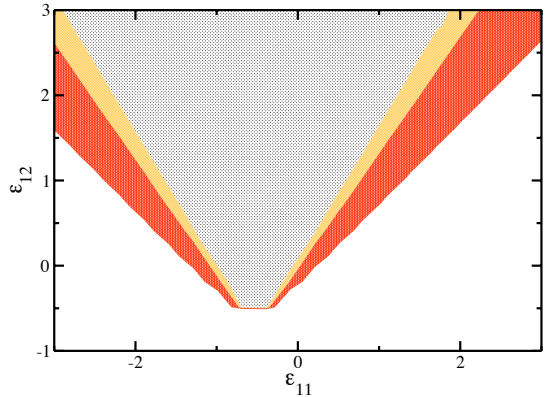}
\caption{The 90 confidence regions for $\theta_{13}$ = 0.0 (the innermost shaded area), 0.1(wider shaded area and 0.2 (widest shaded area). In making this plot $\delta m_{21}^2$ is taken to be $7.59 \times 10^{-5}$ eV$^2$ and $\tan \theta_{12} = 0.68$.}
\label{Fig:X}
\end{center}
\end{figure}

Even though Eq. (\ref{deviation}) includes both the Standard Model physics and physics beyond the Standard Model, it still is a convenient parametrization of the subdominant matter effects. In Figure \ref{Fig:X}  we show our results. Clearly positive values of $\epsilon_{12}$ will effectively increase coupling between flavors, whereas negative values would decrease it. Since experiments indicate that solar electron neutrino flux is much less than expected, coupling between flavors cannot be significantly decreased. Consequently relatively large negative values of $\epsilon_{12}$ 
($\epsilon_{12} < - \sin 2 \theta_{12}/ (2 B_i E)$, $B_i$ being the value of the parameter of Eq. 
(\ref{4}) where the neutrino is produced) 
are not permitted as seen in Fig. \ref{Fig:X}. On the other hand  positive values of $\epsilon_{12}$, since they increase mixing, are favored\footnote{Of course very large values of $\epsilon_{12}$ will yield an energy-independent survival probability (0.5) in contradiction with the data, comparing low- and high-energy neutrinos.} regardless of the value of $\epsilon_{11}$. One also observes that the allowed parameter space is shifted towards negative values of $\epsilon_{11}$. The sign of $\epsilon_{11}$ can be either positive or negative:  non-zero $\theta_{13}$ contribution is negative whereas Standard Model loop corrections are positive. 
To understand this shift, we consider the matter mixing angle in the presence of these additional terms
\begin{equation}
\label{z1}
\cos 2 \theta_m = 
\frac{\cos 2 \theta_{12} - 2BE (1+\epsilon_{11})}{\sqrt{(2BE(1+\epsilon_{11}) - \cos 2 \theta_{12})^2  
+ (2BE \epsilon_{12} + \sin 2 \theta_{12})^2}}
\end{equation}
and its derivative 
\begin{equation}
\label{z2}
\dot{\theta}_m = \frac{1}{2} \left( \frac{\dot{N}_e}{N_e} \right) \frac{2BE[\epsilon_{12} \cos 2 \theta_{12} 
+ (1+ \epsilon_{11} ) \sin 2 \theta_{12}]}{[2BE(1+\epsilon_{11})- \cos 2 \theta_{12}]^2+[2BE \epsilon_{12} +
\sin 2 \theta_{12}]^2}
\end{equation}
in the presence of these additional terms. For the values of $\delta m^2_{12} \sim 10^{-5}$ eV$^2$, the neutrino evolution is adiabatic as can be seen by examining the adiabaticity parameter:
\begin{equation}
\label{z3}
\frac{2E}{\delta m_{12}^2} \left( \frac{\dot{N}_e}{N_e} \right) \frac{2BE[\epsilon_{12} \cos 2 \theta_{12} 
+ (1+ \epsilon_{11} ) \sin 2 \theta_{12}]}{\left([2BE(1+\epsilon_{11})- \cos 2 \theta_{12}]^2+[2BE \epsilon_{12} +
\sin 2 \theta_{12}]^2\right)^{3/2}} . 
\end{equation}
For values of $\epsilon_{11}$ such that the condition $2BE (1+\epsilon_{11})  \sim \cos 2 \theta_{12}$ is satisfied,  an MSW resonance is possible. In this case the adiabaticity parameter takes its maximum value of at the resonance: 
\begin{equation}
\label{z4}
\frac{2E}{\delta m_{12}^2} \left( \frac{\dot{N}_e}{N_e} \right) \frac{2BE(1+ \epsilon_{11} ) \sin \theta_{12}}{(2BE \epsilon_{12} +
\sin 2 \theta_{12})^{1/2}} 
\end{equation} 
which is small in the Sun for $\delta m^2_{12} \sim 10^{-5}$ eV$^2$. 
For larger absolute value of $\epsilon_{11}$ (both positive and negative) clearly the adiabaticity parameter of Eq. (\ref{z3}) is even smaller. (This is also true for large values of $\epsilon_{12}$).  Resulting adiabatic neutrino evolution is then controlled by the initial value of the cosine of the matter angle given in Eq. (\ref{z1}). Regardless of the value of 
$\epsilon_{12}$, for large positive values of $\epsilon_{11}$ this angle tends to $\cos 2 \theta_m \sim -1$, whereas for negative values of $\epsilon_{11}$ this cosine may significantly move away from $-1$.  In the absence of the $\epsilon$ parameters $\cos 2 \theta_m$ is already very close to $-1$. Consequently more of the parameter space for the negative values of $\epsilon_{11}$ is available for fitting the data, shifting the allowed region of the $\epsilon_{11}$ towards negative values for a fixed value of 
$\epsilon_{12}$. This is in concordance with the numerical calculations shown in Fig. \ref{Fig:X}. One should point out that values of $\epsilon_{11}$ and $\epsilon_{12}$ beyond a few percent are likely to be unphysical, but we show relatively large values in Fig. \ref{Fig:X} to illustrate that solar neutrino data do not give practical limits. 

\begin{figure}[t]
\begin{center}
\includegraphics[scale=0.3]{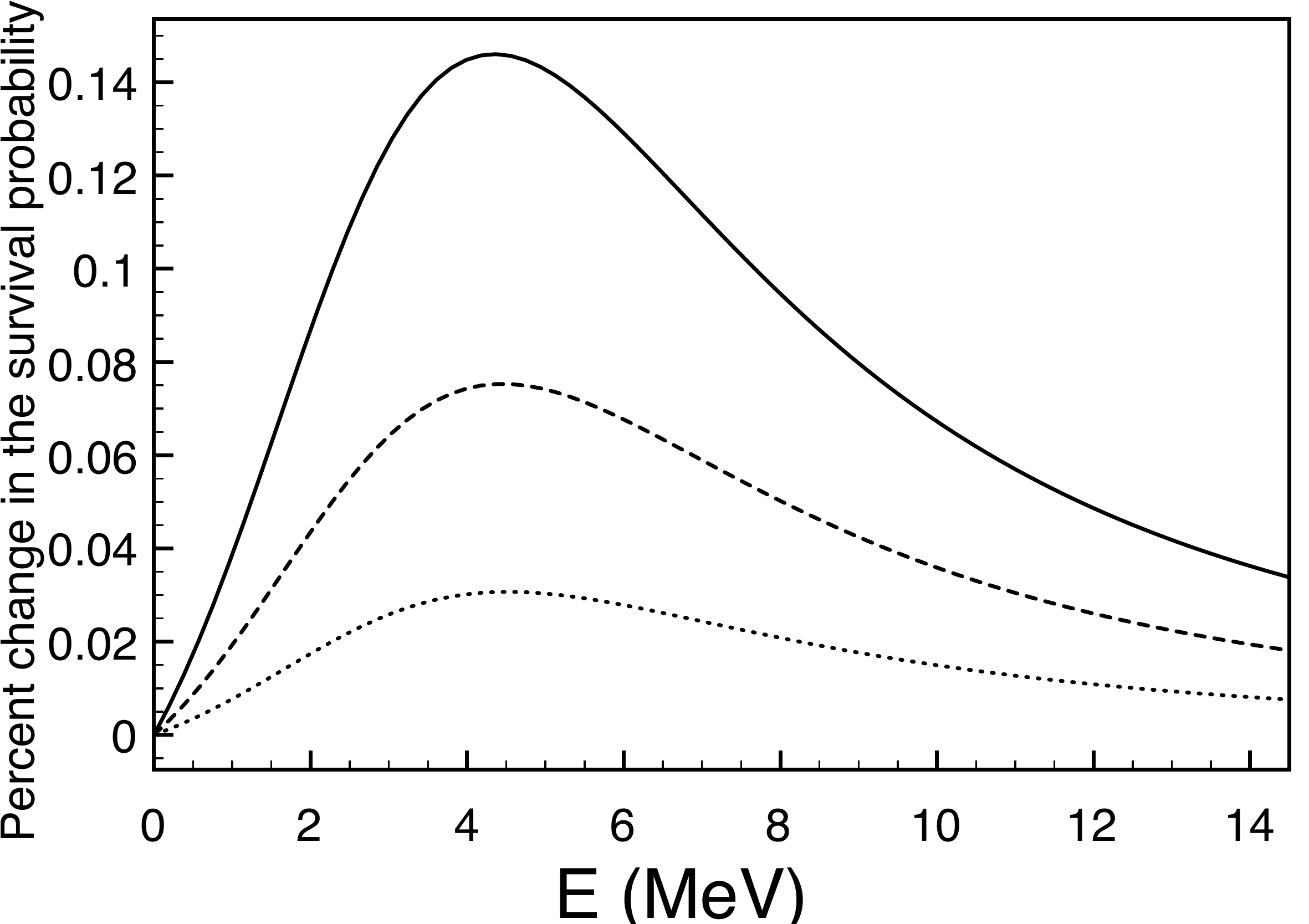}
\caption{Percentage change in the electron neutrino survival probability for $\epsilon_{11} =0.1$ (solid line), $\epsilon_{11} =0.05$ (long-dashed line), and  $\epsilon_{11} =002$ (dotted line). In making this plot $\delta m_{21}^2$ is taken to be $7.59 \times 10^{-5}$ eV$^2$, $\tan \theta_{12} = 0.68$, and 
$\epsilon_{12}=0$.}
\label{Fig:XXX}
\end{center}
\end{figure}

For the smaller (and more realistic) values of the $\epsilon$ parameters the situation is somewhat more optimistic. 
In the LMA region, neutrino propagation in the Sun remains adiabatic if such effects are present and thus the electron neutrino survival probability is controlled by a single parameter, $\cos 2 \theta_i$. The best potential to observe such effects with solar neutrinos then seems to be in the $E_{\nu} \sim $ few MeV region where $2 B_i E$ takes the value $\cos 2 \theta_{12}$, allowing $\epsilon_{11}$ term to be more 'visible'.  We illustrate this behavior in Fig. \ref{Fig:XXX} where the percentage change of the electron neutrino survival probability,
\[
\frac{P_{\nu_e \rightarrow \nu_e}(\epsilon_{11}=0) - P_{\nu_e \rightarrow \nu_e}(\epsilon_{11} \neq 0) }{P_{\nu_e \rightarrow \nu_e}(\epsilon_{11}=0) }, 
\]
is plotted. Future experiments such as the SNO+ experiment \cite{Kraus:2010zz}, currently under construction, should be able to probe such solar neutrino energies. 

\section{Conclusions}

Motivated by the recent low-threshold analysis of the Sudbury Neutrino Observatory and asymmetry measurements of the BOREXINO Collaboration of the solar neutrino flux, we reexamined matter effects on the neutrino propagation in the Sun. Our analysis shows that solar neutrino data constrains the mixing angle $\theta_{13}$ poorly. Both the BOREXINO measurement of the day-night asymmetry of the solar {\em neutrinos} and the KamLAND measurement of the reactor {\em antineutrinos} point out to the LMA region of the neutrino parameter space. In this region solar electron neutrino survival probability is not very sensitive to the value of $\theta_{13}$: to obtain a given value of this probability a wide range of values of $\theta_{13}$ can be used, but permitted variations of $\theta_{12}$ are limited. 

We next revisited the effects of the physics beyond the Standard Model on the neutrino propagation through the Sun. We showed that subdominant Standard Model effects (such as inclusion of a non-zero value of $\theta_{13}$ or loop corrections) can mimic the effects of the physics beyond the Standard Model.  For example, Standard Model loop corrections are very small, but it is not clear that they are smaller than contributions from new physics. We find that , at present,solar neutrino data constrain such effects also rather poorly.

\section*{Acknowledgments}

We thank K. Heeger for useful discussions. 
This work was supported in part
by the U.S. National Science Foundation Grant No. PHY-0855082, and 
in part by the University of Wisconsin Research Committee with funds
granted by the Wisconsin Alumni Research Foundation.

\appendix
\section{Standard Model Corrections to the Two-Flavor Evolution}

Neutrino propagation in matter has been extensively investigated in the context of two-flavor mixing. In this appendix we give the reduction of the full Standard Model contribution (i.e., three active flavors and loop corrections) to two flavors.  We use the following parameterization of the neutrino mixing matrix: 
\begin{equation}
\label{aaa1}
 {\bf T}_{23}{\bf T}_{13}{\bf T}_{12}  = 
\left(
\begin{array}{ccc}
 1 & 0  & 0  \\
  0 & C_{23}   & S_{23}  \\
 0 & -S_{23}  & C_{23}  
\end{array}
\right)
\left(
\begin{array}{ccc}
 C_{13} & 0  & S_{13} e^{-i\delta_{CP}}  \\
 0 & 1  & 0  \\
 - S_{13} e^{i \delta_{CP}} & 0  & C_{13}  
\end{array}
\right) 
\left(
\begin{array}{ccc}
 C_{12} & S_{12}  & 0  \\
 - S_{12} & C_{12}  & 0  \\
0  & 0  & 1  
\end{array}
\right)
\end{equation}
where $C_{ij} = \cos \theta_{ij}$, $S_{ij} = \sin \theta_{ij}$, and $\delta_{CP}$ is the CP-violating phase.  
Following Ref. \cite{Balantekin:1999dx} we introduce the combinations 
\begin{eqnarray}
\label{rot1}
\tilde{\Psi}_{\mu} &=& \cos{\theta_{23}} \Psi_{\mu} -
\sin{\theta_{23}} \Psi_{\tau}, \\
\tilde{\Psi}_{\tau} &=& \sin{\theta_{23}} \Psi_{\mu} +
\cos{\theta_{23}} \Psi_{\tau},
\end{eqnarray}
which yields the MSW evolution equation
\begin{equation} 
\label{CProt}
i \frac{\partial}{\partial t} 
\left(
\begin{array}{c}
  \Psi_e \\
 \tilde{ \Psi}_{\mu} \\
  \tilde{\Psi}_{\tau} 
\end{array}
\right) 
= \tilde{\bf H} 
\left(
\begin{array}{c}
  \Psi_e \\
  \tilde{\Psi}_{\mu} \\
  \tilde{\Psi}_{\tau} 
\end{array}
\right) 
\end{equation}
where 
\begin{equation}
\label{htilde}
\tilde{\bf H} = 
{\bf T}_{13}{\bf T}_{12} 
\left(
\begin{array}{ccc}
E_1  & 0  & 0  \\
0  & E_2  & 0  \\
0  &  0 & E_3  
\end{array}
\right) {\bf T}^{\dagger}_{12}{\bf T}^{\dagger}_{13}  + 
\left(   
\begin{array}{ccc}
 V_{e \mu} & 0  & 0  \\
 0 & S^2_{23}   V_{\tau \mu} & - C_{23} S_{23} V_{\tau \mu}  \\
0  & - C_{23} S_{23}V_{\tau \mu}   & C^2_{23} V_{\tau \mu}  
\end{array}
\right) .
\end{equation}
In writing Eq. (\ref{htilde}), by dropping a term proportional to the identity  $V_{\mu \mu}$ is chosen to be zero.  The non-zero potentials in this equation are 
\begin{equation}
  \label{wolfen1}
V_{\mu e} (x) = \sqrt{2} G_F  N_e (x) 
\end{equation}
and the Standard Model loop correction \cite{Botella:1986wy}, 
\begin{equation}
\label{loop}
V_{\tau \mu} = -  \frac{3 \sqrt{2} G_F\alpha}{\pi \sin^2 \theta_W}  \left( \frac{m_{\tau}}{m_W}\right)^2
\left\{ (N_p + N_n) \log \frac{m_{\tau}}{m_W} + \left( \frac{N_p}{2} + \frac{N_n}{3}  \right) \right\} ,
\end{equation}
which is quite small for the solar densities. 
Note that the initial conditions on $\tilde{\Psi}_{\mu}$ and $\tilde{\Psi}_{\tau}$
are the same as those on $\Psi_{\mu}$ and $ \Psi_{\tau}$: They are 
initially all zero.  We next perform the transformation
\begin{equation}
\label{rot2}
 \left(\begin{matrix}  \varphi_e\cr
    \varphi_{\mu} \cr \varphi_{\tau} \end{matrix}\right) = T_{13}^{\dagger}
 \left(\begin{matrix} \Psi_e\cr
    \tilde{\Psi}_{\mu} \cr \tilde{\Psi}_{\tau} \end{matrix}\right) = 
 \left(\begin{matrix} \cos{\theta_{13}} \Psi_e + \sin{\theta_{13}}
    \tilde{\Psi}_{\tau} \cr
    \tilde{\Psi}_{\mu} \cr - \sin{\theta_{13}}  \Psi_e +
    \cos{\theta_{13}} \tilde{\Psi}_{\tau} \end{matrix}\right),
\end{equation}
after which Eq. (\ref{htilde}) takes the form
\begin{equation}
\label{mswmod2}
i \frac{\partial}{\partial t} \left(\begin{matrix} \varphi_e\cr
    \varphi_{\mu} \cr \varphi_{\tau} \end{matrix}\right) = {\cal H} 
\left(\begin{matrix} \varphi_e\cr
    \varphi_{\mu} \cr \varphi_{\tau} \end{matrix}\right),
\end{equation}
where we dropped a term proportional to the identity. In this
equation ${\cal H}$ is given by 
\begin{equation}
\label{h0new}
{\cal H} = \left(\begin{matrix}
     \frac{1}{2} \tilde{V} - \Delta_{21} \cos 2 \theta_{12} + G &
     \frac{1}{2} \Delta_{21} \sin 2 \theta_{12} +K &  \frac{1}{2} 
     \sin 2 \theta_{13} ( V_{\mu e} - D) e^{-i \delta_{CP}}\ \cr
     \frac{1}{2} \Delta_{21} \sin 2 \theta_{12} + K  & -
     \frac{1}{2}\tilde{V} + \Delta_{21} \cos 2 \theta_{12}  - G  & -\frac{1}{2} C_{13} \sin 2 \theta_{23} V_{\tau \mu} \cr 
     \frac{1}{2} 
     \sin 2 \theta_{13} ( V_{\mu e} - D) e^{i \delta_{CP}}
      & -\frac{1}{2} C_{13} \sin 2 \theta_{23} V_{\tau \mu} 
      &
     \frac{1}{2}(\Delta_{31}+\Delta_{32}) + V_c - \frac{3}{2} 
     \tilde{V} +F \end{matrix} \right),
\end{equation}
where we introduced the modified matter potential 
\begin{equation}
\label{modpotent}
\tilde{V} = V_c\cos^2\theta_{13}, 
\end{equation}
and the quantities 
\begin{equation}
\label{Delta}
\Delta_{ij} = \frac{m_i^2 - m_j^2}{2E} = \frac{\delta m_{ij}^2}{2E}, 
\end{equation}
where
\[
G= V_{\tau \mu} \left( \frac{1}{2} \sin^2 \theta_{13} \cos^2 \theta_{23} - 
     \sin^2 \theta_{23} \right), 
     \]
\[
K = \frac{1}{2} \sin \theta_{13} \sin 2 \theta_{23}  V_{\tau \mu} e^{-i \delta_{CP}} ,
\]
\[
D = \cos^2 \theta_{23} V_{\tau \mu}, 
\]
\[
F = V_{\tau \mu}  \left( \left( 1 - \frac{3}{2} \sin^2 \theta_{13} \right) \cos^2 \theta_{23} - \frac{1}{2} \sin^2 \theta_{23} \right). 
\]

If one sets $V_{\tau \mu} = 0, \delta_{CP}=0$ in Eq. (\ref{h0new}) one gets 
\begin{equation}
\label{h0}
{\cal H} = \left(\begin{matrix}
     \frac{1}{2} \tilde{V} - \Delta_{21} \cos 2 \theta_{12}&
     \frac{1}{2} \Delta_{21} \sin 2 \theta_{12} &  \frac{1}{2} V_c
     \sin 2 \theta_{13} \cr
     \frac{1}{2} \Delta_{21} \sin 2 \theta_{12} & -
     \frac{1}{2}\tilde{V} + \Delta_{21} \cos 2 \theta_{12}   & 0 \cr
      \frac{1}{2} V_c \sin 2 \theta_{13}  & 0 &
     \frac{1}{2}(\Delta_{31}+\Delta_{32}) + V_c - \frac{3}{2} 
     \tilde{V} \end{matrix}\right) .  
\end{equation}
Setting the quantity $\sin 2 \theta_{13}$ in the  off-diagonal terms of Eq. (\ref{h0}) yields the result given in Eq. (\ref{b1}) \cite{Fogli:2001wi,Balantekin:2003dc,Fogli:2008jx,Balantekin:2011ta}.  A similar, but more complicated result can be obtained starting from Eq. (\ref{h0new}). 


\end{document}